\documentclass[11pt,letterpaper]{article}
\usepackage{amsmath, bbm} 
\pdfoutput=1
\usepackage{jheppub} 
\title{Geometric General Solution to the $U(1)$ Anomaly Equations}
\author[a]{B C Allanach,}
\author[b]{Ben Gripaios,}
\author[b]{and Joseph Tooby-Smith\footnote{Corresponding author}}
\affiliation[a]{DAMTP, University of Cambridge, Wilberforce Road, Cambridge, 
CB3 0WA, United Kingdom}
\affiliation[b]{Cavendish Laboratory, University of Cambridge, J.J. Thomson
 Avenue, Cambridge, CB3 0HE, United Kingdom} 
\emailAdd{B.C.Allanach@damtp.cam.ac.uk}
\emailAdd{gripaios@hep.phy.cam.ac.uk}
\emailAdd{jss85@cam.ac.uk}

\abstract{Costa et al.\ [Phys. Rev. Lett. 123, 151601 (2019)] 
recently gave a general
 solution to the anomaly equations for $n$ charges in a $U(1)$ gauge
 theory. `Primitive' solutions of chiral fermion charges were parameterised and it was shown how
 operations performed upon them (concatenation with other
 primitive solutions and
 with vector-like solutions) yield the general solution. 
 We
 show that the ingenious methods used there have a simple geometric
 interpretation, corresponding to elementary constructions in number
 theory. Viewing them in this context allows the fully general solution
 to be written down directly, without the need for further operations. Our geometric method also allows us to show that the only operation 
 Costa et al.\ require is permutation.
 It also 
 gives a variety of other, qualitatively similar, 
 parameterisations of the
 general solution, as well as a qualitatively different (and arguably
 simpler) form of the general solution for $n$ even.}
\newcommand{\Z}{\mathbb{Z}}
\newcommand{\Q}{\mathbb{Q}}

\begin{document}
\maketitle

\section{Introduction}
The local anomaly cancellation equations for a $U(1)$ gauge theory
with $n$ left-handed chiral fermions of charge $z_i$, which may be
 taken to be integers, are
\begin{align}\label{anomalyeq}
\sum_{i=1}^n z_i&=0,\\ \sum_{i=1}^n z_i^3&=0. \label{u1cub}
\end{align}
The first of these, (\ref{anomalyeq}), comes from a one-loop triangle diagram with two external gravitons and one external $U(1)$ gauge
boson~\cite{Eguchi:1980jx}, whilst (\ref{u1cub}) comes from the
similar diagram with three external $U(1)$ gauge
bosons~\cite{Adler:1969gk,Bardeen:1969md,Bouchiat:1972iq,Gross:1972pv,Georgi:1972bb}.
Although written for left-handed chiral fermions, these equations are
general for a theory with both left-handed and right-handed chiral
fermions since we can charge conjugate any right-handed
representation, reversing the sign of its charge and giving a
left-handed representation. Eq.~(\ref{u1cub}) is a
 cubic diophantine equation in $n$ variables; since it is not yet
 known how to solve a generic such equation even in 2 variables
 (corresponding to an elliptic curve~\cite{Hardy}), 
 one might
 expect that finding the general solution to (\ref{anomalyeq}-\ref{u1cub}) is a
 difficult problem. However,
a recent paper by Costa, Dobrescu and Fox
(CDF)~\cite{Costa_Dobrescu_Fox_2019} managed to do so, in the
 following way.

CDF observed that given two integer solutions $\underline{x}:=(x_1,\ldots,x_n)$ and
$\underline{y}:=(y_1,\ldots,y_n)$,
of (\ref{anomalyeq}), and (\ref{u1cub}), a third could be constructed
from a `merger' operation, which they denoted `$\oplus$' 
\begin{align}
 \underline{x} \oplus \underline{y}:=
 \left( \sum_{i=1}^n x_i y_i^2 \right) \underline{x} -
 \left( \sum_{i=1}^n x_i^2 y_i \right) \underline{y}.
 \label{merger}
 \end{align}

Some solutions to (\ref{anomalyeq}) and (\ref{u1cub}) are easy to
find, having for each charge $z_i$ another charge $z_j=-z_i$. Using
solutions of this form, which we call vector-like solutions, and the
merger CDF showed that one can construct chiral sets of charges,
namely those where $z_i+z_j\ne 0$ for all $i$ and $j$. They then
showed  (via rather lengthy algebra) that any solution can be
constructed from these chiral sets of charges by permutation of
charges or concatenation with each other or
with vector-like solutions. For $n$ even the specific mergers they considered were
\begin{align}
 (l_1,k_1, \ldots,k_m,-l_1,-k_1,\ldots,-k_m) \oplus
 (0,0, l_1, \ldots, l_m, -l_1, \ldots, -l_m ),
 \end{align}
 where $m=n/2-1\geq 2$ and
$k_i, l_i \in \Z,\ i \in \{1,\ldots, m\}$. Whilst for $n$ odd they were 
\begin{align}
 (0,k_1, \ldots,k_{m+1},-k_1,\ldots,-k_{m+1}) \oplus
 (l_1, \ldots, l_m,k_1,0,-l_1, \ldots, -l_m,-k_1 ), \label{CDFodd}
\end{align}
where $m=(n-3)/2 \geq 1$. CDF showed that if one wants to avoid zero charges or vector-like copies of charges then conditions have to be applied to $k_i$'s and $l_i$'s.

Here, we show that the ingenious methods of CDF have a simple
geometric interpretation, 
corresponding to elementary constructions long
known to number theorists~\cite{Mordell_1969}.
Viewing them in this context allows a fully general solution
to be written
down in one 
fell swoop.
The geometric interpretation allows us to give a variety of other,
qualitatively similar,
parameterisations of the general solution, as well as a qualitatively
different form of the general solution for even $n$. It also allows us
 to show that to generate all solutions from CDF's parameterisation
 only requires permutations and not the other operations.
 
The paper proceeds as follows: in \S\ref{sec:geom}, we review the geometric
method that we employ to solve (\ref{anomalyeq}) and (\ref{u1cub}),
generalising a number-theoretic result of
Mordell to dimensions higher than 3 in the process.
We
detail two solutions that our method yields directly, but which
require permutations of CDF's solutions, and show that for CDF's parameterisation permutations is the only operation required.
We conclude in \S\ref{sec:dis}. There is one potential inconvenience
in our parameterisation, in that there are special solutions generated differently from others, which we circumvent in 
Appendix~\ref{sec:perms}. 
We present the different form of the general solution for even $n$ in Appendix~\ref{sec:nevenapp}.  

\section{Geometric Method}\label{sec:geom}
By way of motivation, consider the $n=6$ solution $(0,-9,7,-1,8,-5)$ to (\ref{anomalyeq}),
and (\ref{u1cub}). The only way to get this solution using the method
outlined in CDF is by permutation. Our
geometric solution will, on the other hand, be able to generate such
solutions without resorting to permutations.\footnote{Though, as we indicate, utilising permutations can be useful.} The reasoning behind this, as we shall see later, lies in
our use of a geometrical approach, namely that of projective
 geometry over the field $\Q$ of rational numbers. Before seeing
 how geometry makes an appearance in the problem at hand, let us recall the basic definitions.

For a field $k$, the projective space $\mathrm{P}k^{n-1}$ is the space
of all lines through the origin in the affine space $k^n$. In other
words, it is $(k^n-\{0\})/\sim$ where $\sim$ is the equivalence
relation $m_1\sim m_2$ with $m_1,m_2\in k^n$ if and only if there exists a $\lambda \in k$ such that $m_1=\lambda m_2$. We denote a point in $\mathrm{P}k^{n-1}$ by the equivalence classes $[a_1,\cdots,a_n]$ for $a_i\in k$.

Within the projective space $\mathrm{P}k^{n-1}$ we can define $d$-planes. By a $d$-plane (for $d<n-1$) we mean a $d$-dimensional projective subspace of $\mathrm{P}k^{n-1}$, which can be written as
\begin{align}
\Gamma=\sum_{i=1}^{d+1} \alpha_i p_i,
\end{align}
where $[\alpha_1:\cdots:\alpha_{d+1}]\in
\mathrm{P}\mathbb{Q}^{d}$ parameterise the $d$-plane and $p_i\in
\mathrm{P}\mathbb{Q}^{n-1}$ are fixed. A $1$-plane, for example,
is just a (projective) line, homeomorphic to a circle.

To motivate the use of projective space on physical grounds, we note
that the Lie algebra of the $U(1)$ gauge group is isomorphic
to~$\mathbb{R}$. Given that $U(1)$ is compact, this implies that our
charges $z_i$ are not only real-valued, but also commensurate, meaning that if
$z_j\ne 0$, then $z_i/z_j$ is rational for all $i$. We can scale every $z_i$
by a single real parameter without changing the physics, as long
 as the coupling constant is also appropriately scaled. This, along
 with the fact that the $z_i$'s are commensurate, allows us to
 undertake a scaling such that all charges are rational, \emph{viz.}
 $z_i\in\mathbb{Q}$.\footnote{In the end, we can scale them all so
   they are integer, as we previously claimed. But working with the
   field $\mathbb{Q}$, rather than the ring $\mathbb{Z}$, allows us to
   do geometry.} It also tells us that we should think of
 the set of all charges as living in projective space, specifically
 $\mathrm{P}\mathbb{Q}^{n-1}$ and indeed, (\ref{anomalyeq}), and (\ref{u1cub}),
 being homogeneous, define loci therein.

It is convenient for us to eliminate $z_n$ in our equations from the cubic equation in (\ref{u1cub}) to get
\begin{align} \label{cubichyper}
\sum_{i=1}^{n-1} z_i^3-\left(\sum_{i=1}^{n-1} z_i\right)^3=0.
\end{align} 
This equation is homogenous, meaning it is well defined on our
equivalence classes in $\mathrm{P}\mathbb{Q}^{n-2}$, and as such it defines a cubic hypersurface (given it is co-dimension $1$) of $\mathrm{P}\mathbb{Q}^{n-2}$. In order to make progress in solving this equation, we review some geometric methods used in diophantine analysis.  

\subsection{The method of chords}
Consider a homogenous cubic in
  $n$-variables, with rational coefficients, defining a locus in
  $\mathbb{Q}^n$. Let $a$ and $b$ be two points in $\mathbb{Q}^n$ on
  the locus. A result from antiquity\footnote{The result
 certainly goes back at least to Fermat and Newton in the 17th century and
 may go back even further to
 Diophantus in the 3rd century. A historical account is given in
 \cite{Stillwell}.} tells us that a chord between $a$ and $b$ will
intersect the surface at a third point in $\mathbb{Q}^n$. One can
understand this result as follows, let $L(t)=a+t(b-a)$ be the chord
joining $a$ and $b$. Points both lying on this chord and in the cubic
surface must satisfy the equation $kt(t-1)(t-t_0)=0$ where $k,
t_0\in\mathbb{Q}$. This result comes from considering the cubic along
the chord and noting that a cubic has one or three (possibly
degenerate) real roots. Hence within $\mathbb{Q}^n$, there is a third
point of intersection, corresponding to $t=t_0$ and given by
$L(t_0)$. We note that this result is equally valid in projective space,
$\mathrm{P}\mathbb{Q}^{n}$. We will call this construction the
`method of chords'.

Further, a rather more recent (though
equally elementary) result of Mordell~\cite{Mordell_1969} states that {\em
  all}\/ 
rational points in a cubic surface in $\mathrm{P}\mathbb{Q}^2$ can be
constructed from chords in this way, starting from a projective line, $L$, and a
point, $p\notin L$ that both lie in the surface.
It follows
from the realisation that in fact {\em any}\/ point in
$\mathrm{P}\mathbb{Q}^2$ ({\em ergo}\/ any point on the cubic) is on a chord from $p_1$ to a point in
$L$. As we will see, this result generalises in a straightforward way
to $\mathrm{P}\mathbb{Q}^{n}$, but there is no analogous result in affine
space. In $\Q^3$ for example, the analogous result would have to involve two skew lines, $L_1$ and $L_2$. However, points forming a plane with $L_2$ which is parallel to $L_1$ will be missed. In $\mathrm{P}\mathbb{Q}^2$, there is no concept of parallel lines -- pairs of lines are either
disjoint or intersecting -- and indeed the aforementioned points all lie on a chord connecting a point on $L$ to $p$.

This simple observation, when generalised to higher $n$, underlies the
fact that the point $(0,-9,7,-1,8,-5)$ is missing from CDFs $n=6$ parameterisation, but is included when we work in projective space, as we will discuss in detail in \S\ref{sec:compar}.

Before actually using any of these results, we note that our
  general method will not work in the cases for $n=1$, and $n=2$. This
  is because for $n=1$ and $n=2$ it would require a notion of
    a $(-1)$-plane! Part of the discussion, namely that in
  Appendix~\ref{sec:perms}, is also valid only for $n\geq 4$. Happily, the solutions to the $n=1,2,3$ cases can be found directly, allowing us to restrict our general discussion to $n\ge 4$. Namely for $n=1$ the solution is $z_1=0$. For $n=2$,
(\ref{cubichyper}) results in no effective constraint (one obtains that the
left-hand side is identically zero for any $z_1$)
and so the solution of (\ref{anomalyeq}),(\ref{u1cub})
is the point $[z_1:z_2]=[1:-1]\in \mathrm{P}\mathbb{Q}$. We have three solutions for $n=3$: $[1:0:-1]$,
$[0:1:-1]$ and
$[1:-1:0]$. Eqs. (\ref{anomalyeq}) and (\ref{u1cub}) are invariant under
permutations of the
$z_i$
and so these three solutions are all in one equivalence class under such permutations.

We now consider higher $n$ where the results above are more
useful. For illustrative purposes, we
will start with a rather explicit discussion of the case $n=4$. 

\subsection{Application for $n=4$}
Let us consider the cubic anomaly-free
surface in $\mathrm{P}\mathbb{Q}^2$,
\begin{align}
z_1^3+z_2^3+z_3^3-(z_1+z_2+z_3)^3=0, \label{neq4}
\end{align}
corresponding to the $n=4$ case of our problem, where we remember that
$z_4=-(z_1+z_2+z_3)$ from the gravitational mixed anomaly constraint. Using Mordell's result within this surface we take the line
$\Gamma_1=[k_1:k_2:-k_1]$ and the point $\Gamma_2=[0:l_1:-l_1]$ in
$\mathrm{P}\mathbb{Q}^2$, which are easily seen to lie on the cubic. Using the overall scaling of projective space, we
could rescale such that $l_1=1$.
At this stage, however, we refrain from doing so, preferring a slightly
redundant parameterisation in order to stay closer to our analysis
of the higher $n$ cases below. We then construct a line passing through
a generic point
on each of
$\Gamma_1$ and $\Gamma_2$ as
$L_1=\alpha_1[k_1:k_2,:-k_1]+\alpha_2[0:l_1:-l_1]$, where
$k_{1,2}, l_1 \in \Q$.
The homogeneous
parameter $[ \alpha_1 : \alpha_2 ] \in \mathrm{P} \Q^1$
parameterises $L_1$, which
must intersect the cubic
surface at a third point, assuming that $L_1$ is not wholly in the cubic
surface. On substituting the 
chord
into (\ref{neq4}) we obtain the constraint on $\alpha_1$ and $\alpha_2$ at
intersections of the line and the cubic surface:
\begin{align}
-3 (k_1 - k_2) l_1 \alpha_1 \alpha_2 \left[(k_1 + k_2) \alpha_1 + 
 l_1 \alpha_2\right]=0. \label{eq:squarebrk}
\end{align}
If $L_1$ were entirely in the cubic surface, the left-hand side would have
evaluated to zero independently of the values of $k_1, k_2$ or $l_1$. 
The third point of intersection is specified by
setting the square bracket in (\ref{eq:squarebrk}) to zero, i.e.\
\begin{align}[\alpha_1:\alpha_2]=[l_1:-(k_1+k_2)],\label{rat}\end{align}
a rational point.

Now consider an arbitrary point $[a_1:a_2:a_3]\in \mathrm{P}\mathbb{Q}^2$ 
{\em not}\/ in $\Gamma_2$. We can define a line between this point and one on
$\Gamma_2$: $L_2=\beta_1[0:l_1:-l_1]+\beta_2[a_1:a_2:a_3]$. It can be seen that this line
intersects $\Gamma_1$ at $[\beta_1:\beta_2]=[a_3-a_1:l_1]$. This, combined with (\ref{rat}), tells us that every such
rational solution to the cubic equation can be found by considering lines
between points on $\Gamma_1$ and $\Gamma_2$. What we have done here is apply Mordell's result to solve the $n=4$ case of our problem. 
\subsection{Arbitrary $n \geq 4$}
To consider arbitrary values of $n \geq 4$ we must generalise Mordell's result to an
arbitrary cubic 
{\em hyper}surface $X$ in $\mathrm{P}\mathbb{Q}^{n-2}$. The generalisation is immediate and gives the following
\newline
{\bf Theorem}: Let $\Gamma_1, \Gamma_2\subset X$ be disjoint
planes of dimensions
$d_1,d_2=m_o:= (n-3)/2$, if $n$ is odd and of dimensions
$d_1=m_e := (n-2)/2$ and $d_2=m_e-1$ if $n$ is even. Every rational
point $p\in \mathrm{P}\mathbb{Q}^{n-2}$ ({\em ergo}\/ every $p \in X$) lies on a chord joining a point
in $\Gamma_1$ to a point in $\Gamma_2$. 
\newline
{\bf Proof}: The result is obvious if $p \in \Gamma_2$. If $p\notin
\Gamma_2$, then $p$ and $\Gamma_2$ define a ($d_2+1$)-plane,
which intersects $\Gamma_1$ in a point $p^1$. The line through $p$ and
$p^1$ intersects $\Gamma_2$ in a point $p^2$, yielding a chord. QED.

In the case of interest,
the (projective) line $L=\alpha_1 p^1+\alpha_2 p^2$ through $p^{1,2}$ with homogeneous parameter
$[\alpha_1:\alpha_2] \in \mathrm{P}\mathbb{Q}^1$ 
intersects the cubic hypersurface $X$ defined by (\ref{cubichyper}) when
\begin{align}
3\alpha_1 \alpha_2\sum_{i=1}^{n-1}\left(\alpha_1 p^2_i P^1_i+\alpha_2 p^1_i P^2_i\right)=0,\;
 P^a_i:= (p^a_i)^2-\left(\sum_{j=1}^{n-1} p^a_j\right)^2. \nonumber
\end{align}
Thus, along with the points $p^{1,2}$ (corresponding to
$\alpha_{2,1}=0$) we get either a third rational point on $X$ at
\begin{align}
[\alpha_1:\alpha_2]=\Bigg[\sum_{i=1}^{n-1}p^1_i P^2_i:-\sum_{i=1}^{n-1}p^2_iP^1_i\Bigg], \label{ourParam}
\end{align}
or, if the terms on the right-hand side both vanish, we have that every rational point on
$L$ is on $X$. Lines which lie in $X$ may be regarded as slightly
awkward to deal with. Happily, it is possible, as
we show in Appendix~\ref{sec:perms}, to find every solution on such a line by a
permutation of the coordinates of a solution arising as the unique third point of intersection on a line not lying in $X$. A comparison of (\ref{ourParam}) with
(\ref{merger}) shows that the `merger'
operation is really nothing but the finding of the third rational point starting from two
others.

To get a fully general solution, we just need to find suitable
$\Gamma_1,\Gamma_2$. To wit,
\begin{align} 
\Gamma^{e}_{1}&=[k_1: \cdots: k_{m_e}:k_{m_e+1}:-k_{1}:\cdots:-k_{m_e}]\nonumber \\
\Gamma^e_{2}&=[0:l_1: \cdots: l_{m_e}:-l_1:\cdots:-l_{m_e}] \nonumber \\
\Gamma^o_1 &=[k_1:\cdots:k_{m_o+1}:-k_1:\cdots:- k_{m_o+1}]\nonumber \\
\Gamma^o_{2}&=[l_2:\cdots: l_{m_o}:l_{m_o+1}:0:-l_1:\cdots:-l_{m_o}:-l_{m_o+1}]. \label{specificHP}
\end{align}
These planes are disjoint (only meeting at the origin, which is not in $\mathrm{P}\mathbb{Q}^{n-2}$), so by the
Theorem they yield all rational
solutions of (\ref{anomalyeq}). 

\subsection{Comparison with CDF} \label{sec:compar}
The parameterisations of CDF, in contrast to ours, have $k_{m_e+1}=-l_1$
and $l_{m_o+1}=k_1$. We have already discussed above that CDF's
solution misses the point $(0,-9,7,-1,8,-5)$, for $n=6$ and that for
them this has to be found by permuting another solution, for example
that generated with $k_1=14$, $k_2=2$, $l_1=-18$, $l_2=-9$ after
scaling. In our parameterisation $(0,-9,7,-1,8,-5)$ can be
obtained directly with, for example, $k_3=0$, $k_1=3$, $k_2=-2$, $l_1=1$, and $l_2=-1$ in (\ref{specificHP}), giving $p^1=[3,-2,0,-3,2]$ and $p^2=[0: 1: -1: -1: 1]$ and the correct third point of intersection.

It is easy to see why CDF's parameterisation misses this point; they
cannot set both $k_3$ and $l_3$ to zero. Viewing things in the affine space $\mathbb{Q}^5$, the geometric nature of such missed points becomes manifest. The planes for $n=6$ in (\ref{specificHP}) can be seen as corresponding to
\begin{align}
\tilde \Gamma_1^e&=(k_1,k_2,1,-k_1,-k_2)\\
\tilde \Gamma_2^e&=(0,l_1,l_2,-l_1,l_2).
\end{align}
in $\mathbb{Q}^5$. The $3-d$ plane defined by $\tilde \Gamma_2^e$ and the point $(-9,7,-1,8,-5)$ does not intercept the $2-d$ plane $\tilde \Gamma_1^e$, which is the same reason why Mordell's result fails to catch all the points in $\Q^3$. CDF go halfway to allowing such points, but by fixing $k_3=l_1$ they don't quite catch them all.

We can be more specific and ask: given the planes in (\ref{specificHP}) where we force $k_{m_e+1}=-l_1$ and $l_{m_0+1}=k_1$ to retrieve CDF's solution, what points don't lie on lines between them? It is easy to see that for even $n$ this would require either $k_{m_e+1}$ or $l_1$ to be zero and for odd $n$ either $l_{m_0+1}$ or $k_1$, but not both. Thus, for the point $[a_1:\cdots:a_n]$ to not lie on such a line, we need, for even $n$,
\begin{align}
a_1+\cdots+a_{n-1}=0\text{~or~}a_1+a_{n-2}=0,
\end{align} 
or, for odd $n$,
\begin{align}
a_1+\cdots+a_{n-1}=0\text{~or~}a_{m_0+2}=0.
\end{align} 
For a non-zero solution we can always rearrange the charges so that none of these conditions are satisfied.

The only other points CDF miss are those where the line between the
two planes in (\ref{specificHP}) lies within $X$. For example for
$n=4$ setting $k_2=k_1$ gives a line $L$ which lies in $X$. As an
explicit example, consider $k_{1,2},l_1=1$. This line is given by
\begin{align}
L=\alpha_1[1:1:-1]+\alpha_2[0:1:-1].
\end{align}
For CDF, points on this line correspond to solutions of the form $(A,-A,B,-B)$ for $A,B\in \mathbb{Z}$. However, CDF's $n=4$ parameterisation
\begin{align}
(-l_1^3(k_1+l_1),-k_1l_1^2(k_1+l_1),k_1l_1^2(k_1+l_1),l_1^3(k_1+l_1))
\end{align}
can never land on such solutions. Nevertheless, CDF's parameterisation
can get these by permutations, for the same reason that the
parameterisation given here can, as we discuss in Appendix~\ref{sec:perms}. 

The above two points not only show when CDF's parameterisation fails to reach a specific point but also proves that their parameterisation produces every point up to permutations.

\section{Discussion} \label{sec:dis}
The pioneering work of CDF finds solutions to
the local $U(1)$ anomaly cancellation constraints. This allows the
construction of the general solution, provided one allows permutations. Our geometric 
method provides the general solution directly without having to perform 
additional steps. The geometric method also explains how some of the otherwise
obscure features of CDF's construction (particularly the `merging' procedure
of two solutions) come about. Due to an immediate generalisation of a theorem by Mordell, the geometric
method is guaranteed to find {\em all}\/ rational solutions for a fixed number
of charges $n$. Therefore (after clearing all denominators), it finds all
integer solutions.

Two further remarks are in order. Firstly, as we have seen, our parameterisation of the
general solution is somewhat distasteful, in that occasionally the
chord $L$ joining points on $\Gamma_{1,2}$ lies in $X$, and so yields not
one, but infinitely many solutions.
Another way to find these solutions is to permute the coordinates $z_i$ of
solutions arising as the unique third intersection of a line $L$ which is
not in $X$, 
as shown in Appendix~\ref{sec:perms}.
Secondly, in the case where $n$ is even, a completely
different, and arguably even simpler, construction of a general solution is
possible. Indeed, in 
such cases, the cubic hypersurface has double points, where both the
left-hand side of (\ref{cubichyper}) and its partial derivatives vanish
(e.g.\ the rational point $[+1:-1:+1:-1:\ldots:+1:-1:+1]$). A
line through such a double point intersects the cubic in one other
rational point (or the line lies entirely in $X$) and thus all
solutions can be obtained by constructing all lines through just a single
double point, as it were. This is worked through explicitly in Appendix~\ref{sec:nevenapp}.
\section*{Acknowledgements}
This work was supported by STFC consolidated grants ST/P000681/1 and
ST/S505316/1. We thank other members of the Cambridge Pheno Working Group for
helpful discussions. 

\appendix
\section{Any solution via permutations \label{sec:perms}}
Here, we give a proof of the statement that any solution sitting on a
line in $X$ between the $d$-planes defined in~(\ref{specificHP}) can be found
by the permutation of the coordinates of a solution which is on a line not in $X$. 
The proof of this statement follows similar reasoning
to the proof regarding permutations of solutions
in~\cite{Costa_Dobrescu_Fox_2019}. 
We must distinguish between $n$
odd and even, so we do them each in turn.

\subsection{Even $n\ge 4$}
We redefine variables such that 
\begin{align}
&x_i=z_1, &&\text{for $i=1$},\nonumber\\& x_i=z_i+z_{m_e+i}, &&\text{for $i=2,\cdots,m_e+1$},\nonumber\\ &y_i=z_i+z_{m_e+1+i}, &&\text{for $i=1,\cdots,m_e$}.\nonumber
\end{align}
The $d$-planes in (\ref{specificHP}) are defined in our new variables by $y_i=0$ for
$\Gamma_1^{e}$ and $x_i=0$ for $\Gamma_2^e$. 
Consider a point $p=[x_i:y_i]\notin \Gamma^{e}_1 \cup
\Gamma^{e}_2$. There is a unique line 
\begin{align}
L_p=\alpha p^1+\beta p^2,\nonumber
\end{align}
through $p$,
$p^1\in \Gamma^{e}_1$ and $p^2\in\Gamma^{e}_2$. Under the
permutation $\phi^e: z_{m_e+1}\leftrightarrow z_{2m_e+1}$, only $y_{m_e}$
changes and
\begin{align}
 L_{\phi^e(p)}=\alpha p^1+\beta\phi^e(p^2).\nonumber
\end{align}
A necessary condition for $L_p$ to be in $X$ is that 
\begin{align}
 -3
y_{m_e}x_{m_e+1}\left(2\sum_{i=1}^{m_e}x_i+x_{m_e+1}\right)+\cdots
&=0\Leftrightarrow\nonumber\\
-3(z_{m_e}+z_{2m_e+1})(z_{m_e+1}+z_{2m_e+1})\left(2\sum_{i=1}^{2m_e+1}z_i-z_{m_e+1}-z_{2m_e+1}\right)+\cdots&=0,\nonumber
\end{align}
where the dots indicate terms which are independent of $y_{m_e}$.

Thus if $L_p$ is in $X$, for a solution $p$ with coordinates permuted such that
\begin{align}
|z_{m_e+1}|\ne |z_{2m_e+1}|\quad \text{and}\quad z_{m_e+1}+ z_{2m_e+1}-\sum_{i=1}^{2m_e+1}z_i\ne 0,\nonumber
\end{align}
then $L_{\phi^e(p)}$ will not be in $X$. The only case where this cannot be
done is where all $|z_i|$ are equal, but such solutions already occur in
$\Gamma_1^e$ after permutations of the $z_i$.

\subsection{Odd $n \geq 4$}
Here,
\begin{align}
& x_i=z_{m_0+1},&&\text{for $i=1$},\nonumber\\& x_i=z_{i-1}+z_{m_0+1+i},&&\text{for $i=2,\cdots,m_o+1$},\nonumber\\ &y_i=z_i+z_{m_o+1+i},&&\text{for $i=1,\cdots,m_o+1$}.\nonumber
\end{align} 
Again,
$\Gamma^{o}_1$ is simply defined by $y_i=0$ and $\Gamma^{o}_2$ is defined
by $x_i=0$. Similar to the even $n$ case, we take a point $p=[x_i:y_i]\notin \Gamma^{o}_1 \cup
\Gamma^{o}_2$. There is again a unique line 
\begin{align}
L_p=\alpha p^1+\beta p^2,\nonumber
\end{align}
through $p$,
where $p^1\in \Gamma^o_1$ and $p^2\in\Gamma^{o}_2$.
Taking
$\phi^o:z_1\leftrightarrow z_{m_o+2}$, only $x_2$ changes, where
\begin{align}
L_{\phi^o(p)}=\alpha \phi^o(p^1)+\beta p^2.\nonumber
\end{align}
A necessary condition for $L_p$ to be in $X$ is then
\begin{align}
-3 x_2y_1\left(2\sum_{i=2}^{m_o+1}z_i+y_1\right)+\cdots
&=0\Leftrightarrow\nonumber\\ -3(z_1+z_{m_o+3})(z_1+z_{m_o+2})\left(2\sum_{i=1}^{2m_0+2}z_i-z_1-z_{m_0+2}\right)
+ \cdots &=0,\nonumber
\end{align}
where now the dots indicate terms which are independent $x_2$.

If $L_p$ is in $X$ for a solution $p$ with coordinates permuted such that 
\begin{align}
|z_1|\ne |z_{n_o+2}|\quad \text{and}\quad z_1+z_{m_o+2}-2 \sum^{2m_o+2}_{i=1}z_i\ne 0,\nonumber
\end{align}
then $L_{\phi^o(p)}$ will not be in $X$.
We may use this construction for all solutions and $n$ odd. 

\section{Alternative solution for $n$-even} \label{sec:nevenapp}
For even $n$, the cubic equation in (\ref{cubichyper}) has double points; that
is points where all of the partial derivatives of the left-hand side 
vanish, as well as the left-hand side itself. An example of such a double point is
\begin{align}
d=[+1:-1:+1:-1:\ldots:+1:-1:+1]\in \mathrm{P}\mathbb{Q}^{n-2}.
\end{align}
So for \emph{e.g.} $n=6$, we have $[+1:-1:+1:-1:+1]$. 

Consider a line through our double point $d$, $L=\gamma_1 d+\gamma_2 r$, for
$r\in\mathrm{P}\mathbb{Q}^{n-2}$ a fixed point and $[\gamma_1:\gamma_2]$
specifying the position along the line. Any point in
$\mathrm{P}\mathbb{Q}^{n-2}$ lies on such a line, and further every such line
is either in the hypersurface $X$ (defined by (\ref{cubichyper})) or passes
through that hypersurface at exactly one other point.  

This other point of intersection can be found by substituting $L$ into (\ref{cubichyper}):
\begin{align}\label{eq:nevensub}
\gamma_2^2\left(3\gamma_1\sum_{i=1}^{n-1}d_iR_i +\gamma_2\sum_{i=1}^{n-1}r_i R_i \right)=0,\quad R_i:=r_i^2-\left(\sum_{j=1}^{n-1} r_j\right)^2.
\end{align}
Either $\gamma_2=0$ (the original point $d$), the LHS is zero independently of $\gamma_1$ and $\gamma_2$ (corresponding to $L$ being in $X$) or 
\begin{align}
[\gamma_1:\gamma_2]=\left[\sum_{i=1}^{n-1}r_i R_i:-3\sum_{i=1}^{n-1}d_iR_i\right],
\end{align}
giving the second point of intersection. As such we can see that the lines $L$
can be used to find all solutions to (\ref{cubichyper}) parameterised by
$r_i$, and if $L$ is in $X$ by $[\gamma_1:\gamma_2]$. 

Continuing our example, for $n=6$, we have that (\ref{eq:nevensub}) becomes
\begin{align}
3\gamma_1(r_1^2-r_2^2+r_3^2-r_4^2+r_5^2-(r_1+r_2+r_3+r_4+r_5)^2)\nonumber \\+\gamma_2(r_1^3+r_2^3+r_3^3+r_4^3+r_5^3-(r_1+r_2+r_3+r_4+r_5)^3)=0
\end{align}
implying the second point of intersection is at
\begin{align}
[\gamma_1:\gamma_2]=&[(r_1^3+r_2^3+r_3^3+r_4^3+r_5^3-(r_1+r_2+r_3+r_4+r_5)^3) \nonumber
\\ &:-3(r_1^2-r_2^2+r_3^2-r_4^2+r_5^2-(r_1+r_2+r_3+r_4+r_5)^2)].
\end{align}

\bibliographystyle{JHEP-2}
\bibliography{comment}

\end{document}